\documentclass[fleqn,10pt]{wlscirep}
\usepackage[utf8]{inputenc}
\usepackage[T1]{fontenc}

\usepackage{ulem}

\usepackage{etoolbox}

\title{A dataset of primary nasopharyngeal carcinoma MRI with multi-modalities segmentation}

\author[1,$\dag$]{Yin Li }
\author[2,$\dag$]{Qi Chen}
\author[1,$\dag$]{Meige Li}
\author[3]{Liping Si}
\author[1]{Yingwei Guo}
\author[1]{Yu Xiong}
\author[1]{Qixing Wang}
\author[1]{Yang Qin}
\author[2]{Ling Xu}
\author[4,5]{Patrick van der Smagt}
\author[1,*]{Kai Wang}
\author[1,*]{Jun Tang}
\author[5]{Nutan Chen}
\affil[1]{Department of Otorhinolaryngology, The First People's Hospital of Foshan, China}
\affil[2]{Department of Radiology, The Second Affiliated Hospital of Anhui Medical University, China}
\affil[3]{Department of Radiology, Zhongshan Hospital, Fudan University, Shanghai, China}
\affil[4]{ELTE University, Budapest, Hungary}
\affil[5]{Foundation Robotics Labs, Munich, Germany}

\affil[*]{corresponding author(s): 
Jun Tang (fsyyytj@126.com) 
Kai Wang (wangkai.sysu@163.com)}

\affil[$\dag$]{these authors contributed equally to this work}

\begin{abstract}
Multi-modality magnetic resonance imaging(MRI) data facilitate the early diagnosis, tumor segmentation, and disease staging in the management of nasopharyngeal carcinoma (NPC).
The lack of publicly available, comprehensive datasets limits advancements in diagnosis, treatment planning, and the development of machine learning algorithms for NPC.
Addressing this critical need, we introduce the first comprehensive NPC MRI dataset, encompassing MR axial imaging of 277 primary NPC patients.
This dataset includes T1-weighted, T2-weighted, and contrast-enhanced T1-weighted sequences, totaling 831 scans. In addition to the corresponding clinical data, manually annotated and labeled segmentations by experienced radiologists offer high-quality data resources from untreated primary NPC.

\end{abstract}
\begin{document}

\flushbottom
\maketitle

\thispagestyle{empty}

\section*{Background \& Summary}
Nasopharyngeal carcinoma (NPC) is the sixth high-incidence cancer globally, with an estimated age-standardized incidence rate of 1.5 per 100,000 person-years as of 2020\cite{zhang2023nasopharyngeal}. However, it exhibits significant geographical variations, with much higher incidence rates in Southeast Asia, where it can reach up to 7.7 and 2.5 per 100,000 for males and females, respectively \cite{zhang2023nasopharyngeal}. Epstein-Barr virus (EBV) infection, genetic factors, family history, and environmental factors are attributed to the risk of incidence\cite{su2024epidemiology}. Men have three times higher morbidity than women, in addition to a peak incidence typically between 40 and 60\cite{koide2024definitive}. The five-year survival rate for NPC can range from around 60\% to 90\%, depending on the stage of the cancer at diagnosis, 14\% of them suffered local recurrence, while 21\% had concomitant distant metastases\cite{au2018treatment}. 
Radiotherapy is a primary treatment for NPC, especially for early-stage tumors. Induction chemotherapy (ICT) combined with concurrent chemoradiotherapy (CCRT) or CCRT alone is the treatment regimen of patients with advanced NPC\cite{petit2023role}.
Intensity-modulated radiation therapy (IMRT) is the preferred method due to its ability to target the tumor more precisely and reduce radiation-related toxicities to surrounding healthy tissues; thereby the contouring of primary gross tumor volume (GTVp) plays a crucial role regarding multimodal or multiparametric imaging data before radiotherapy \cite{chua2016nasopharyngeal}.
Follow-up is recommended by the Chinese Society of Clinical Oncology and European Society for Medical Oncology guidelines, consisting of a combination of periodic clinical, 
laboratory and radiological examinations including nasal endoscope, magnetic resonance imaging, and plasma EBV-DNA concentration\cite{tang2021chinese,bossi2021nasopharyngeal}.
MRI is recommended to be performed every 6–12 months for surveillance of the local disease and the late complications\cite{tang2021chinese}.

MRI is superior to CT in presenting soft-tissue contrast, multi-modality imaging, and no ionizing radiation. CT excels in visualizing bony changes in the skull base and adjacent intracranial and extracranial structures, particularly in assessing foramina, fissures, and other intricate bony anatomy of the skull base. Positron emission tomography (PET) plays a vital role in the metabolic assessment of NPC. It is particularly effective in detecting distant metastases, evaluating treatment response\cite{wei2016comparison}. MRI, on the other hand, offers several advantages that make it the preferred modality for comprehensive NPC assessment: 1. Multi-modality imaging: MRI provides multiple contrast mechanisms (T1-weighted, T2-weighted, and enhanced-contrast sequences) that offer diverse tissue characterization. 2. Multiplanar capabilities: MRI allows for high-resolution imaging in multiple planes (axial, coronal, and sagittal), facilitating a comprehensive 3D understanding of tumor extent. 3. Superior soft tissue resolution: MRI clearly differentiates between muscles, mucosa, and adipose tissue, enabling precise delineation of tumor margins and invasion patterns, and retropharyngeal lymph node\cite{widmann2017mri}. 4. Enhanced tumor staging accuracy: The high soft tissue contrast of MRI allows for precise delineation of tumor invasion, which is critical for accurate staging. MRI also has demonstrated superior accuracy in detecting and re-staging the recurrence or residual primary lesions\cite{thamboo2022surveillance}. Furthermore, MRI is significantly more cost-effective than PET/CT, making it a more accessible and practical imaging modality for routine surveillance and local assessment in NPC patients. 5. Early detection of submucosal lesions: MRI can reveal subtle submucosal abnormalities that may not be apparent on CT or endoscopic examination.

Given these capabilities, MRI is widely regarded as the imaging modality of choice for the localization, characterization, and staging of nasopharyngeal lesions. Its superior soft tissue contrast and multiparametric capabilities make it an essential tool in the diagnostic workup and treatment planning for patients with suspected or confirmed NPC.

The computer-assisted delineation of GTVp in NPC based on multi-modality MRI and CT imaging data has been widely used by oncologists during pre-radiotherapy planning. \cite {lin2019deep,peng2023improved,chen2020mmfnet}. Annotation of NPC imaging by an experienced radiologist is a time-consuming procedure, and its individual-dependent variability may compromise the accuracy of IMRT.

As the application of machine learning (ML), 
especially convolutional neural networks (CNNs) \cite{lecun1998gradient,krizhevsky2012imagenet}, U-Net \cite{ronneberger2015u}, and vision Transformers (ViTs) \cite{dosovitskiy2020image}
In medical image analysis, automatic segmentation, disease prediction, and evaluation have advanced significantly.  Annotated multi-sequence MRI data combined with clinical parameters, provide multidimensional features to train and validate models \cite{li2022npcnet,huang2019achieving,chen2020mmfnet,lin2019deep,zhong2021deep,kayalibay2017cnn}.
Recently, many curated automatic segmentation models have been proposed to offer the delineation of GTVp for NPC\cite{tang2021dsunet,hao2023msu,gu2023cddsa,luo2023deep}. The  extracted features from the manually segmented images in the field of deep learning offer highly accurate ground truth, more detailed annotation, and individual subtlety correction which heavily rely on experienced oncologists and radiologists. The challenge of obtaining high-quality annotations, combined with the prevalent absence of publicly available datasets in methodological studies, highlights the critical need for enhanced data sharing and the establishment of open-access imaging repositories to support robust AI development.
The labor-intensive specificity of accurately annotating medical images requires significant expertise and resources, making it a bottleneck in the development cycle of ML applications in medical imaging. 
By making these datasets publicly available, researchers and developers can leverage multidimensional features to improve model performance, innovate in automated segmentation, and enhance disease diagnosis and treatment strategies. Furthermore, published datasets enable the scientific community to collaborate more efficiently, replicate and validate findings, and accelerate the progress in medical imaging technologies of GTVp for NPC.

We presented a dataset of 277 patients with a clinical and pathological diagnosis of NPC with corresponding MR imaging axial slices including three sequences: T1-weighted, T2-weighted, and contrast-enhanced T1-weighted. The data included the laboratory examination regarding EBV status, biopsy results, and progression-free survival in the five-year follow-up period.

\section*{Methods}
\begin{figure}[!ht]
    \centering
    \includegraphics[width=0.6\textwidth]{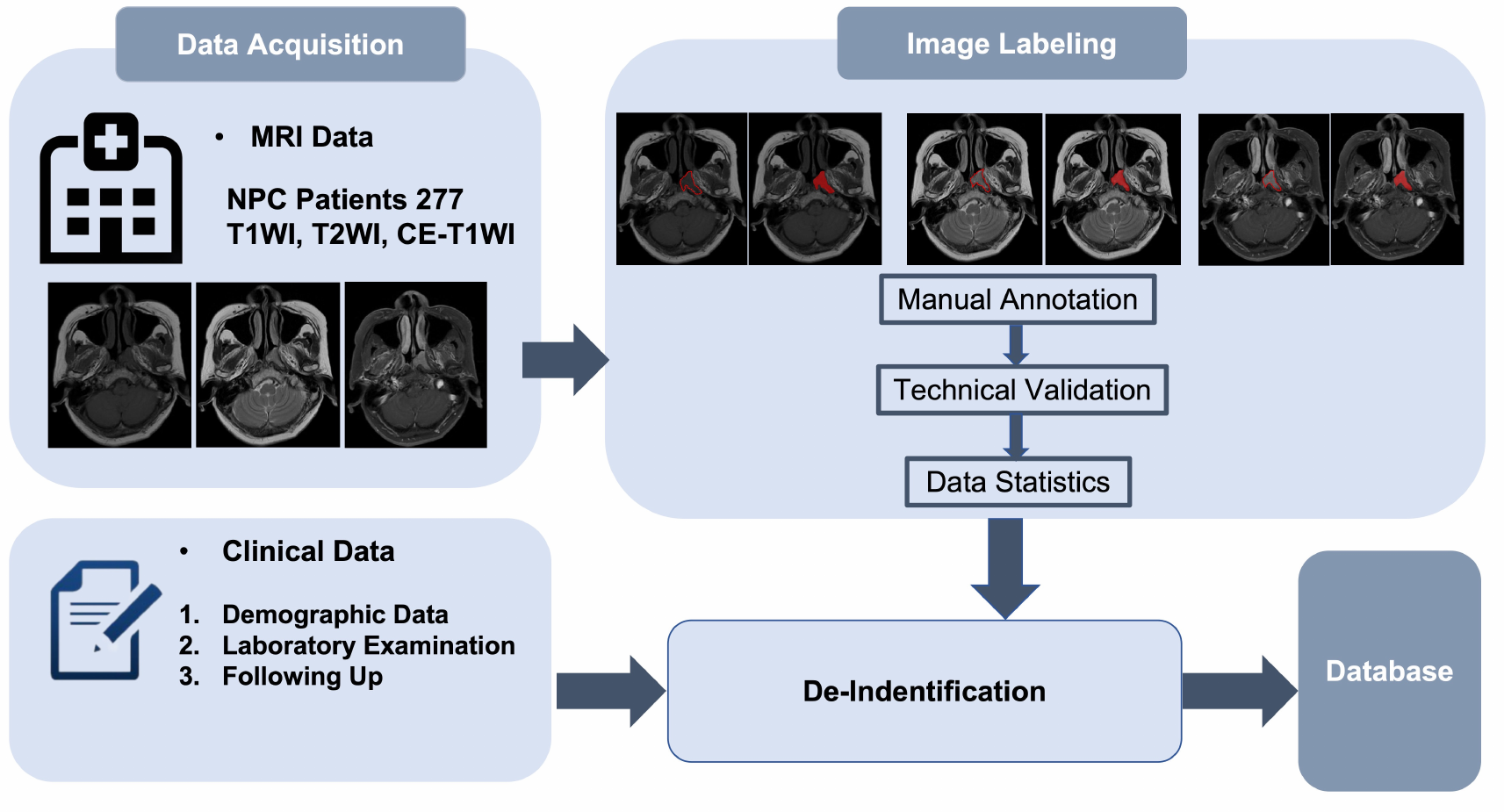}
    \caption{Graphic abstract and showcase. 
    The MRI showcases a patient's MRI images including T1WI, T2WI, and CE-T1WI sequences, which were processed by the manual annotation with labeled tumor boundaries and highlighted areas, respectively.
}
    \label{fig:abs}
\end{figure}

\begin{table}[ht]
\centering
\begin{tabular}{|l|c|c|c|}
\hline
& T1WI & T2WI & CE-T1WI
\\
\hline
Echo time (TE) (ms) & $12.0 \,(8.6 - 16.4)$ & $80.0 \,(66.46 - 123.09)$ & $12.0 \,(1.67 - 16.4)$ \\
\hline
Repetition time (TR) (ms) & $570.88 \,(366.67 - 2283.34)$ & $3192.87 \,(2500.0 - 7633.34)$ & $632.0 \,(330.0 - 1100.0)$ \\
\hline
Spacing between slices (mm) & $5.5 \,(0.5 - 7.0)$ & $5.5 \,(0.5 - 7.0)$ & $5.5 \,(0.5 - 7.0)$ \\
\hline
Slice thickness (mm) & $5.0 \,(4.0 - 6.2)$ & $5.0 \,(4.0 - 6.2)$ & $5.0 \,(4.0 - 6.2)$ \\
\hline
Pixel spacing (mm) & $0.43 \,(0.38 - 0.55)$ & $0.43 \,(0.38 - 0.98)$ & $0.45 \,(0.4 - 0.55)$ \\
\hline
Field of view (FoV) ($\mathrm{mm}^2$) & $220\times 220 \,(220 - 250)$ & $225 \times 225 \,(220 - 250)$ & $220 \times 220 \,(220 - 250)$ \\
\hline 
Flip angle (degrees) & $90.0 \,(65.0 - 142.0)$ & $90.0 \,(90.0 - 180.0)$ & $90.0 \,(80.0 - 142.0)$ \\
\hline
Matrix size & $512 \,(512 - 640)$ & $512 \,(256 - 640)$ & $512 \,(432 - 640)$ \\
\hline
\end{tabular}
\caption{MRI parameters of 277 patients. The results are represented as median (min-max) for all parameters.}
\label{tab:parameters}
\end{table}

\begin{table}[ht]
\centering
\begin{tabular}{|l|c|c|c|}
\hline
& T1WI & T2WI & CE-T1WI \\
\hline
Volume ($\mathrm{cm^3}$) & $8.60 \,(0.68 - 70.34)$ & $8.67 \,(0.40 - 73.12)$ & $9.42 \,(0.61 - 84.24)$ \\
\hline
Surface area ($\mathrm{cm^2}$) & $38.03 \,(6.25 - 198.49)$ & $37.12 \,(4.55 - 199.80)$ & $40.35 \,(5.38 - 230.52)$ \\
\hline
Max diameter ($\mathrm{cm}$) & $4.51 \,(2.14 - 9.96)$ & $4.51 \,(1.73 - 10.08)$ & $4.70 \,(2.00 - 10.15)$ \\
\hline
Surface regularity & $0.40 \,(0.23 - 0.63)$ & $0.40 \,(0.23 - 0.62)$ & $0.39 \,(0.23 - 0.84)$ \\
\hline
\end{tabular}
\caption{Tumor size of 277 patients. The results are represented as median (min-max).}
\label{tab:volume}
\end{table}

\begin{table}[ht]
\centering
\begin{tabular}{|l|c|c|c|c|c|c|}
\hline
machine & Achieva & DISCOVERY MR750w & MAGNETOM IMPACT & OPTIMA MR360 & SIGNA EXCITE & Signa HDxt \\
\hline
amount & 146 &  12 &  93 &   5 &   1 &  20 \\
\hline
\end{tabular}
\caption{Machines}
\label{tab:machines}
\end{table}

Our cohort included a total of 277 patients diagnosed with primary nasopharyngeal carcinoma in The First People's Hospital of Foshan, China. All selected subjects were histopathologically confirmed as primary NPC without a prior history of radiation therapy, chemotherapy, or other malignancies that could potentially distort the tumor morphology. All patients were restaged according to the eighth edition of the Union for International Cancer Control/American Joint Committee on Cancer (UICC/AJCC) staging system\cite{brierley2017tnm}. The collected patients with age, gender, TNM stages, histopathological diagnosis, EBV status (VCA-IgA, EBV-DNA) and the 5-year progression-free survival further contributed to the clinical database. The process of data is depicted in Figure \ref{fig:abs}.

In our study, the image acquisition process was meticulously structured to ensure the highest quality of data for investigating nasopharyngeal carcinoma (NPC) segmentation. 
All cases underwent both non-contrast and contrast-enhanced MRI scans, providing comprehensive imaging and clinical data. The exclusion criteria included: 1. patients who had undergone radiotherapy or chemotherapy before the MRI examination, as the internal tumor structure and lesion boundaries may change after treatment, preventing an accurate reflection of the tumor's original growth state; 2. patients with a history of other malignant tumors, which may confound variables and affect the study’s outcomes;
3. images that did not meet quality standards: 
(1) Incomplete scan coverage: When lesion areas were excessively large, parts of the lesions may have extended beyond the imaging field, resulting in incomplete evaluation of the entire NPC lesion.
(2)	Insufficient image resolution: Low-resolution images hindered accurate segmentation of tumor boundaries, potentially leading to imprecise measurements and analysis.
(3)	Presence of artifacts: Various imaging artifacts,such as motion or susceptibility effects, could obscure or distort the appearance of lesions, reducing  the reliability of assessments.

The dataset was captured using six MR scanners from manufacturers including GE Discovery MR750w 3.0T and Philips Achieva 1.5T systems with Gadoteric Acid Meglumine Salt as the contrast agent.
The setup and calibration of the six MR scanners (see Tab. \ref{tab:machines}) were similar, as can be seen from the STD of machine parameters in Table \ref{tab:parameters}.

To ensure the privacy and confidentiality of patient information, a strict anonymization protocol was applied. 
Personal identifiers such as names, ages, birth data, sex, weight, content date were removed. Furthermore, unique patient IDs were replaced with the patients' index within the dataset. Indirect identifiers were also deleted to prevent any potential re-identification, e.g., study date and institute information.

%%%%%%
\begin{table}[htbp]
\centering
\begin{tabular}{|l|c|c|c|c|c|}
\hline
\textbf{Clinical Stage} & \textbf{1 (N=5)} & \textbf{2 (N=26)} & \textbf{3 (N=140)} & \textbf{4a (N=94)} & \textbf{4b (N=13)} \\ \hline
\textbf{Gender}         &                  &                   &                    &                    &                    \\ \hline
F                       & 1 (20.0\%)       & 10 (38.5\%)       & 41 (29.3\%)        & 18 (19.1\%)        & 4 (30.8\%)         \\ \hline
M                       & 4 (80.0\%)       & 16 (61.5\%)       & 99 (70.7\%)        & 76 (80.9\%)        & 9 (69.2\%)         \\ \hline
\textbf{Age}            &                  &                   &                    &                    &                    \\ \hline
Mean (SD)               & 51.8 (11.4)      & 49.7 (13.5)       & 50.8 (13.0)        & 49.5 (12.6)        & 58.0 (9.86)        \\ \hline
Median                  & 45.0             & 48.5              & 50.0               & 49.0               & 59.0               \\ \hline
Min, Max                & [44, 70]         & [27, 80]          & [22, 91]           & [20, 74]           & [36, 71]           \\ \hline
\textbf{Pathological Type} &              &                   &                    &                    &                    \\ \hline
Basaloid Squamous Cell Carcinoma & 0 (0\%)  & 0 (0\%)           & 0 (0\%)            & 0 (0\%)            & 1 (7.7\%)          \\ \hline
Keratinized Squamous Cell Carcinoma & 1 (20.0\%) & 2 (7.7\%) & 5 (3.6\%)        & 4 (4.3\%)          & 0 (0\%)            \\ \hline
Non-Keratinized Squamous Cell Carcinoma & 4 (80.0\%) & 24 (92.3\%) & 135 (96.4\%)  & 90 (95.7\%)        & 12 (92.3\%)        \\ \hline
\textbf{T Stage}        &                  &                   &                    &                    &                    \\ \hline
T0                      & 0 (0\%)          & 0 (0\%)           & 0 (0\%)            & 0 (0\%)            & 1 (7.7\%)          \\ \hline
T1                      & 5 (100.0\%)      & 8 (30.8\%)        & 2 (1.4\%)          & 0 (0\%)            & 0 (0\%)            \\ \hline
T2                      & 0 (0\%)          & 18 (69.2\%)       & 41 (29.3\%)        & 3 (3.2\%)          & 2 (15.4\%)         \\ \hline
T3                      & 0 (0\%)          & 0 (0\%)           & 97 (69.3\%)        & 15 (16.0\%)        & 4 (30.8\%)         \\ \hline
T4                      & 0 (0\%)          & 0 (0\%)           & 0 (0\%)            & 76 (80.9\%)        & 6 (46.2\%)         \\ \hline
\textbf{N Stage}        &                  &                   &                    &                    &                    \\ \hline
N0                      & 4 (80.0\%)       & 6 (23.1\%)        & 10 (7.1\%)         & 5 (5.3\%)          & 0 (0\%)            \\ \hline
N1                      & 1 (20.0\%)       & 20 (76.9\%)       & 31 (22.1\%)        & 31 (31.9\%)        & 1 (7.7\%)          \\ \hline
N2                      & 0 (0\%)          & 0 (0\%)           & 99 (70.7\%)        & 29 (30.9\%)        & 4 (30.8\%)         \\ \hline
N3                      & 0 (0\%)          & 0 (0\%)           & 0 (0\%)            & 30 (31.9\%)        & 8 (61.5\%)         \\ \hline
\textbf{M Stage}        &                  &                   &                    &                    &                    \\ \hline
M0                      & 5 (100.0\%)      & 26 (100.0\%)      & 140 (100.0\%)      & 93 (98.9\%)        & 0 (0\%)            \\ \hline
M1                      & 0 (0\%)          & 0 (0\%)           & 0 (0\%)            & 1 (1.1\%)          & 13 (100.0\%)       \\ \hline
\textbf{EBV}            &                  &                   &                    &                    &                    \\ \hline
Negative                & 5 (100.0\%)      & 21 (80.8\%)       & 109 (77.9\%)       & 58 (61.7\%)        & 7 (53.8\%)         \\ \hline
Positive                & 0 (0\%)          & 5 (19.2\%)        & 31 (22.1\%)        & 36 (38.3\%)        & 6 (46.2\%)         \\ \hline
\textbf{VCA}            &                  &                   &                    &                    &                    \\ \hline
Negative                & 4 (80.0\%)       & 6 (23.1\%)        & 45 (32.1\%)        & 19 (20.2\%)        & 3 (23.1\%)         \\ \hline
Positive                & 1 (20.0\%)       & 20 (76.9\%)       & 95 (67.9\%)        & 75 (79.8\%)        & 10 (76.9\%)        \\ \hline
\textbf{Overall Survival} &               &                   &                    &                    &                    \\ \hline
Mean (SD)               & 48.0 (26.8)      & 49.7 (22.3)       & 52.8 (18.5)        & 45.7 (21.0)        & 22.2 (15.2)        \\ \hline
Median                  & 60.0             & 60.0              & 60.0               & 60.0               & 18.0               \\ \hline
Min, Max                & [0, 60.0]        & [0, 60.0]         & [0, 60.0]          & [0, 60.0]          & [5.0, 60.0]        \\ \hline
\end{tabular}
\caption{Clinical Stage Data}
\end{table}

\subsection*{Standardization and Calibration of MRI Machines}

To ensure consistency across different imaging sessions, a rigorous standardization and calibration protocol was implemented for all MRI scanners following international quality assurance standards. The calibration process aimed to minimize variations due to machine-specific differences and environmental factors.

\subsubsection*{Calibration Procedure}

Each MRI machine was calibrated before the commencement of the study and checked regularly throughout the imaging period. The main aspects of the calibration process included:

\begin{itemize}
  \item \textbf{Geometric calibration:} Ensuring the geometric accuracy of the images by calibrating the scanner’s spatial resolution settings using a standardized phantom object.
  \item \textbf{Signal intensity calibration:} Standardizing the signal intensity levels by adjusting the MR signal parameters to match a predefined baseline, ensuring consistent image brightness and contrast across sessions.
  \item \textbf{Magnetic field homogeneity adjustment:} Regularly assessing and optimizing the magnetic field homogeneity to reduce artifacts and improve the accuracy of the image data.
\end{itemize}

\subsubsection*{Quality Assurance}

To maintain the calibration standards, quality assurance tests were conducted periodically. These tests involved:

\begin{itemize}
  \item Acquiring images of a standard test phantom that includes structures to evaluate resolution, contrast, and signal uniformity.
  \item Comparing these images against reference images to detect any deviations that might indicate a need for recalibration.
\end{itemize}

\subsection*{Segmentation procedure}
Recognizing the importance of high-quality manually segmented data in advancing NPC imaging research, we have decided to share our meticulously curated dataset of manual NPC segmentations. This dataset, created by experienced radiologists using a standardized protocol, encompasses multiple MRI sequences and provides a valuable resource for researchers and developers working on improving NPC segmentation techniques.

The segmentation procedure was a critical step in preparing MRI data for effective analysis. Since the manual segmentation is widely regarded as the gold standard due to its high accuracy, all the MRI images in this study were manually segmented.

1. Image Review and Tumor Identification:
All three sequence images were thoroughly reviewed to define tumor regions and areas of surrounding invasion. The process involved:
(1) Localization: Confirming that the lesion originated from the nasopharyngeal mucosa or submucosa, typically presenting as early-stage mucosal thickening or soft tissue mass formation. On T1WI, lesions appeared as equal or high signals, while on T2WI, they showed high signals. CE-T1W images demonstrated significant enhancement.
(2) Invasion Assessment: Larger lesions often exhibited invasion into surrounding structures. This could manifest as:
   a. Parapharyngeal space involvement, characterized by the disappearance of surrounding fat planes and muscle invasion.
   b. Upward extension with skull base bone destruction and intracranial invasion. Bone destruction was evident as high-signal yellow bone marrow was replaced by low-signal tumor tissue.
   c. Intracranial invasion commonly affects the cavernous sinus, temporal lobe, and cerebellopontine angle region.
   d. Cervical lymph node metastases, typically showing a top todown, ipsilateral to contralateral sequential pattern.

2. Tumor Boundary Determination:
Nasopharyngeal MRI scans usually include T1, T2, and CE-T1 sequences in axial, coronal, and sagittal planes. 
Radiologists integrated information from all three sequences to accurately determine tumor boundaries:
    T1WI effectively displayed the surrounding fat spaces and muscle structures.
   CE-T1WI was crucial for precise boundary determination of early mucosal thickening.
 T2WI helped differentiate tumor mass from mucosa, with the mass typically showing lower signal intensity than the mucosa.

3. Manual Segmentation Procedure:
Two experienced diagnostic radiologists, each with over ten years of work experience, independently performed layer-by-layer manual segmentation of lesion boundaries using ITK-SNAP 3.6.1 software (version 3.6.1).
\cite{py06nimg}
%(\url{http://www.itksnap.org}). 
The polygon mode drawing icon was utilized for all three sequences in the axial plane.

4. Specific Steps for Manual Segmentation:
(1) Data Import: 
   - DICOM format neck MR images exported from the Picture Archiving and Communication System (PACS), including the nasopharyngeal region, were imported into ITK-SNAP software.
(2) Data Retrieval and Visualization:
   Window width and level were adjusted for optimal lesion edge visualization. Relevant slices containing lesions were selected for ROI annotation.
(3) ROI Delineation and Adjustment:
   ROIs were manually delineated layer-by-layer, following the inner lesion boundary to reduce partial volume effects. For extensive tumors, care was taken to exclude adjacent structures (e.g., blood vessels, lymph nodes) while including direct invasion areas. Difficult boundaries in one sequence were cross-referenced with other sequences. Both continuous and point-by-point delineation methods were used, with manual adjustments as needed to ensure accuracy. After completing all layers, 3D segmentation images were generated for each sequence per patient.

The segmented tumor regions were then converted into binary masks, representing the presence or absence of tumor tissue in each pixel.  This rigorous segmentation procedure ensured the creation of a highly accurate and reliable dataset, forming a robust foundation for our subsequent analyses on NPC segmentation.

\subsection*{Technical Validation}
\begin{figure}[!ht]
    \centering
    \includegraphics[width=0.6\textwidth]{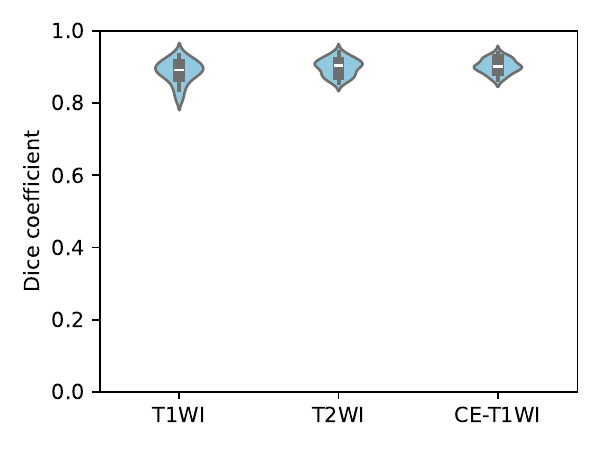}
    \caption{Inter reliability. Comparison of Dice coefficients between two radiologists across 30 patients.
}
    \label{fig:dice}
\end{figure}

\begin{figure}[!ht]
    \centering
    \includegraphics[width=0.6\textwidth]{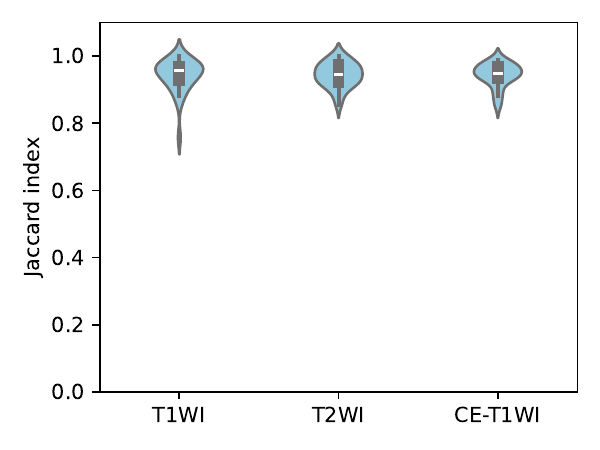}
    \caption{Inter reliability. Comparison of Jaccard index between two radiologists across 30 patients.
}
    \label{fig:jaccard}
\end{figure}

To perform technical validation, we used Inter-Rater Reliability on a randomly selected set of 30 patient images. Two experienced radiologists, each with over 10 years of expertise, independently delineated the tumor lesions on three MRI sequences for each patient. The inter-rater reliability was assessed by comparing the Dice coefficients and Jaccard indices between the two radiologists across all 30 patients. The results, presented in Fig.~\ref{fig:dice} and Fig.~\ref{fig:jaccard}, demonstrated high consistency between the radiologists, confirming the reliability of the manual delineation process.
Validation upon completion, the senior radiologist randomly selected and reviewed 10 cases to ensure compliance; all of them matched segmentation standards.

\subsection*{Morphological Parameters}

In the analysis of 3D tumor models, the study utilizes critical metrics such as Surface Area, Volume, Max Diameter, and Surface Regularity to understand the spatial attributes of tumors, particularly those with irregular shapes or multi-directional extensions.

\begin{itemize}
    \item \textbf{Surface Area Measurement:} The Marching Cubes algorithm constructs a triangular mesh of the tumor surface. The total surface area is then calculated by summing the areas of all the triangles in the mesh.
    \item \textbf{Volume Calculation:}

    Based on the mesh, the method calculates the contribution of each triangle to the total volume. For each triangle, the algorithm computes a signed volume, taking into account the triangle’s orientation, which is determined by the order of its vertices. This approach effectively handles the irregular borders of the tumor by accurately incorporating the contribution of each individual triangle in the mesh.
    \item \textbf{Max Diameter:} This metric is computed by identifying the maximum Euclidean distance between any two points on the tumor surface mesh.
    \item 
    \textbf{Surface Regularity:}
    Surface Regularity
$= v / (\frac{4}{3} \pi R^3)$ 
where $v$ represents the volume of the segmented tumor and $R$ denotes the equivalent radius of the tumor computed by the surface area.
Surface Regularity is particularly important in understanding the complexity of a tumor's shape, as tumors with more irregular surfaces can indicate different biological behaviors compared to those with smoother surfaces.
\end{itemize}

These metrics provide a comprehensive assessment of the tumor's characteristics, aiding in diagnosis and treatment planning.

\subsection*{Method comparison}
Manual segmentation of NPC is a time-consuming and labor-intensive process. Recent advancements in artificial intelligence and machine learning have led to the development of increasingly accurate automatic and semi-automatic segmentation methods for NPC.
\cite{chen2020mmfnet}
\cite{li2022ddnet}

These studies highlight the growing trend towards more sophisticated and accurate automatic segmentation methods. However, the development and validation of such advanced techniques require high-quality, manually segmented datasets as a foundation for training and evaluation. While automatic segmentation methods continue to evolve and improve in accuracy, manually segmented data remains an indispensable component in the development and refinement of these algorithms. Manual segmentations serve as the ground truth for training, validation, and performance assessment of automated methods.

\subsection*{Ethical statement}
All methods of the present research were approved by the  Board Committee on Ethics of the First People's Hospital of Foshan, Foshan, China (Number:2023-96), confirming that all data collection procedures adhered strictly to international guidelines on research ethics, ensuring the privacy and confidentiality of all participants involved. The waiver of consent was granted by the ethnic committee of the First People's hospital of Foshan.

\section*{Data Records}

\begin{figure}[]
    \centering
    \includegraphics[width=0.45\textwidth]{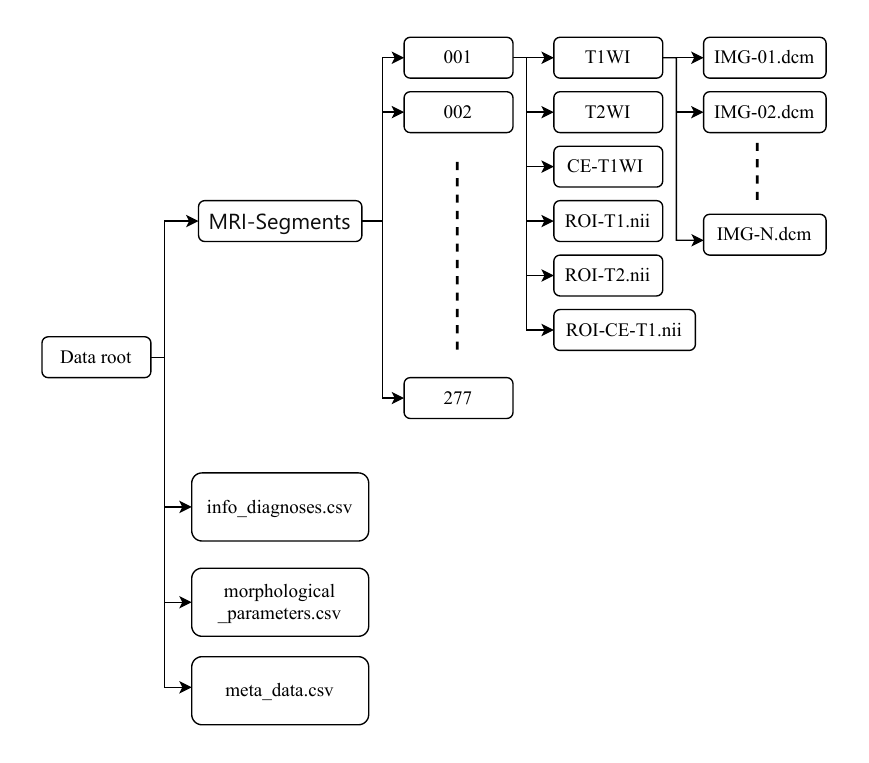}
    \caption{Dataset Structure.
}
    \label{fig:datasturcture}
\end{figure}

The dataset \cite{Li2024NPCdataset} is available on \url{https://zenodo.org/records/10900202}. 
It is distributed under the Creative Commons Attribution 4.0 International License (CC BY 4.0) (\url{https://creativecommons.org/licenses/by/4.0/}). 
The structure of the data in \texttt{primary\_data.zip} is depicted in Figure~\ref{fig:datasturcture}. 
The folder \texttt{MRI-Segments} contains subfolders for each patient with MRI scans and segmentation data. 
The imaging data is stored in DICOM format, which is widely recognized as the standard for storing and exchanging medical images due to its comprehensive metadata capabilities. The segmentations are stored in NIfTI format, which is commonly used in neuroimaging and offers advantages such as ease of use with various analysis tools and efficient storage of 3D data.
The file \texttt{info\_diagnoses.csv} includes diagnostic details, \texttt{morphological\_parameters.csv} contains morphological data, and \texttt{meta\_data.csv} provides information about the MRI machine types.
Additionally, we have another independent file \texttt{inner\_reliability.zip} that includes $256 \times 256$ resolution images and masks. In our toolkit repository, we provide the code to convert the data to customized resolution images. Finally, \texttt{inner\_reliability.zip} contains the labels used to compute inner reliability.

To visualize segmentation, ITK-SNAP
is useful. For reading MRI data in Python, the PyDICOM library
is available. Moreover, vtkmodules and Nibabel are suitable for processing ROI data in Python. These tools have been specifically tested and verified, but other general software and libraries are also capable of reading and visualizing this dataset.

\section*{Code availability}
The acquisition and annotation of data were conducted without the use of code. For the analytical procedures described in this manuscript, the code is accessible at \url{https://github.com/leexxe/NPC-MRISegmentationToolkit.git}.

\section*{Author contributions statement}

Yin Li: Conception and design, Data collection, Data processing, Analysis and interpretation, Writing–original draft, Writing–review and editing. Qi Chen: Data labeling, Data processing, Writing–original draft, Writing–review and editing. Kai Wang:  Data collection, Data processing. Meige Li: Data collection, Data processing. Liping Si: Data labeling. Yingwei Guo: Data collection. Yu Xiong: Data collection. Qixing Wang: Data collection. Yang Qin: Data collection. Ling Xu: Data labeling, Data collection. Patrick van der Smagt: Conception and design, Data collection, Data processing, Writing–original draft. Jun Tang: Conception and design, Data collection, Data processing, Analysis and interpretation, Writing–original draft, Writing–review and editing. Nutan Chen: Conception and design, Data collection, Data processing, Analysis and interpretation, Writing–original draft, Writing–review and editing.

\section*{Competing interests}
The authors declare no competing interests.

\section*{Acknowledgment}
This present study is supported by the Medical Scientific Research Foundation of Guangdong Province, China (B2025811) and the Project of Foshan Science and Technology Bureau (2220001003814).

\bibliography{sample}

\end{document}